 \documentclass[article]{elsarticle}

\usepackage{amssymb}
\usepackage{lipsum}
\usepackage{mathptmx}
\usepackage{etoolbox}
\usepackage{natbib}  
\journal{Sensors and Actuators: A. Physical}

\begin{document}

\begin{frontmatter}



\title{PCB-Integrated CoPt Micromagnets for Magnetophoresis}


\author[first]{Melissa Mitchell}
\affiliation[first]{organization={University of Bath},
            addressline={}, 
            city={Bath},
            postcode={BA2 7AY},
            country={UK}}

\author[first]{Henrique Mira}

\author[first]{Simon Bending}

\author[second]{Chris Bell}
\affiliation[second]{organization={University of Bristol},
            addressline={}, 
            city={Bristol},
            postcode={BS8 1QU}, 
            country={UK}}

\author[first]{Ali Mohammadi}

\begin{abstract}
Integration of magnetic material with scalable microfluidic platforms can significantly improve the throughput and precision in biotechnology processes. In this work we have developed a new magnetic platform based on printed circuit board (PCB) technology. Cobalt-Platinum (CoPt) micromagnets are electroplated on copper pads with an arbitrary footprint on a Kapton substrate to enable generation of different magnetic field gradients. The magnets, with areas ranging 9mm\textsuperscript{2}, 13.5mm\textsuperscript{2}, and 18mm\textsuperscript{2}, are characterized by XRD and VSM, before and after thermal annealing. The ordered L1\textsubscript{0} phase appear in XRD results after annealing at 600 \textdegree C, and VSM results show around six times increase for in-plane magnetic remanence from 0.24T to 1.4T. The near equiatomic ratios of Co:Pt is confirmed by EDX observations. The performance of these magnets is experimentally validated by trapping magnetic nanoparticles in microfluidic channels. These results are in excellent agreement with FEA models presented in COMSOL Multiphysics.
\end{abstract}



\begin{keyword}
PCB \sep Electroplating \sep Cobalt-Platinum \sep Magnetic Nanoparticles \sep Magnetic Trapping \sep Lab-on-PCB



\end{keyword}

\end{frontmatter}




\section{Introduction}
\label{introduction}

The integration of electromagnetic sensors and actuators into printed circuit board (PCB) technology has been well established through the use of planar copper coils, including surface-etched transformers and inductive elements, in applications such as power conversion, motor control, and sensing.\cite{marignetti2017electromagnetic, samuolis2022development, dimarco2019calibration}. However, certain niche applications, including magnetophoresis, require significantly stronger magnetic field gradients, especially focused at small scales e.g., within microfluidic channels, to enable effective trapping of labelled cells \cite{alnaimat2018microfluidics}. The performance of these PCB-compatible planar coils as sources of magnetic fields is constrained by Ohmic losses, and limited flux concentration leading to electromagnetic cross-coupling.
To mitigate such constraints, integration of ferromagnetic materials on PCBs such as soft magnetic composites (SMCs) has been investigated \cite{10861315}. Despite the versatility and design flexibility of SMCs, hard ferromagnetic materials with high energy density are required in magnetophoresis. 
These permanent magnets offer stronger field gradients without any external power source; which is also advantageous for point-of-care devices compared to electromagnetic coils \cite{permanent}. Furthermore, permanent magnets are less influenced by the size downscaling constraints seen in electromagnets; under certain conditions, the field intensities of magnets remain constant under geometric downscaling \cite{magmas}.

Current approaches for integrating hard magnetic materials into PCBs generally involve the manual insertion of sintered blocks, the deposition of bonded magnetic pastes, or electrolytic deposition \cite{9718871}. Although sintered magnets provide high flux density, limited scalability and mechanical discontinuities in the substrate present manufacturing challenges\cite{thomas2020ferrites}. Bonded pastes offer geometric freedom and ease of deposition, but sacrifice magnetic strength due to the need for polymer binders \cite{1427815}. Hence, electroplating remains a viable alternative due to its monolithic integration with various alloys (e.g., CoNi, NiFe), micron-scale geometric resolution and high flux concentration \cite{pattanaik2007electrodeposition}.


For high-energy density applications including magnetophoresis, hard magnets such as CoPt and FePt are promising choices for scalable manufacturing due to compatibility with electrodeposition. Despite the higher energy density of FePt, it is more prone to delamination at thick film scales limiting its versatility when compared to CoPt \cite{thongmee2007fept, https://doi.org/10.1049/mnl.2019.0287, magnetophoresis}.

Further to scalable manufacturing, the size and strength of the magnets are key factors in the precision and accuracy of these devices \cite{water, cervical, heart, rnas}. The magnets need to be strong enough to capture the MNPs and labelled cells, without damaging the samples. The high field strength of traditional bulk magnets can cause paramagnetic beads to cluster, leading to blockages in the microfluidic channels \cite{block1, block2}. Hence, integrating miniaturised magnets with microfluidic platforms can significantly improve the precision of high-throughput detection systems. 

The patterned magnetic platform described by Osman et al. is used to trap cells \cite{4}. However, the manufacturing process for patterning the magnetic material requires complex laser-based magnetic annealing leading to scalability issues. Further to this, magnetic microwells and the flyover technique are presented in Huang et al. \cite{microwell} and Lin et al. \cite{flyover}, respectively, to guide the field lines of bulk external magnets for trapping labelled cells. Despite significant advances, further work is needed to provide a scalable technology for integrated high-precision magnets in microfluidic platforms. 

This paper presents electroplated CoPt micromagnets on a PCB platform, with potential biotechnology applications in magnetophoresis. These low-cost substrates enjoy scalable manufacturing processes such as photolithography and electrodeposition with permanent magnets that offer a higher energy density. The details of process technology are explained in Section II. An FEA model is presented in Section III to explain the operation of magnets for trapping particles. Section IV describes the results achieved by SEM, EDX, XRD, VSM, and proof-of-concept trapping experiments.


\section{Materials and Manufacturing of CoPt Magnets on PCB}





\begin{figure}
\centering
    \includegraphics[width=0.9\linewidth]{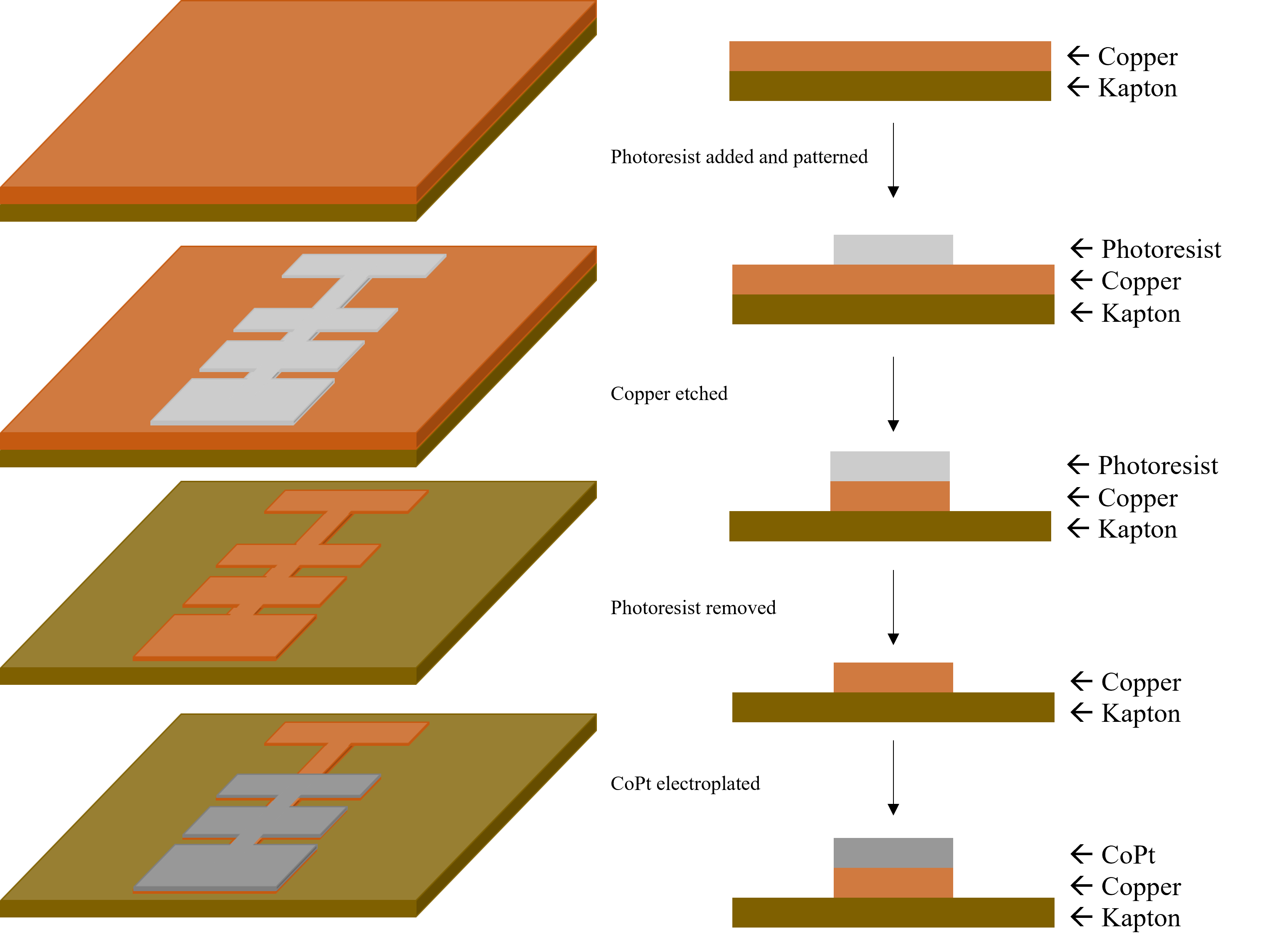}\\
    
    \caption{\label{fig:cross-sec} Proposed process for preparing the substrate, including cross-section illustrations for each step. 3D illustration (left) and cross-section (right).}
    
\end{figure}

Figure \ref{fig:cross-sec} shows the manufacturing steps of the proposed integration technique. The flexible PCB substrate is designed with four rectangular copper pads linked by a connecting copper line. Three of these rectangular pads are used for plating the magnets, and the top pad is used as a contact pad for the cathode during electroplating. The microfabrication steps, including cross-sections, are illustrated in figure 1. The plating pads are all 0.45cm wide, and the lengths are 0.4cm, 0.3cm, or 0.2cm. The width was determined by the width of a bundle of microfluidic channels used for magnetic bead separation in the following steps \cite{ianniello2025dna}. 
\\

The flexible substrate, a copper-on-Kapton sheet, is patterned using Ordyl FP415 dry photoresist, which was laminated onto the substrate. After developing this photoresist, the area where the copper was to remain for the plating pads and the connecting trace are covered with photoresist. This is then submerged in ferric chloride to remove the copper surrounding the pads. This process step etches the copper everywhere except for the plating pads, cathode connector pad, and the trace line as shown in the third step in Figure \ref{fig:cross-sec}. Conventionally, a photoresist mould may be used to ensure electroplating in a specified area. This photoresist must be removed before the high-temperature annealing step, as otherwise results in carbonization and contamination. Removing the surrounding copper means that there is no need for photoresist, making the later annealing step simpler. Furthermore, previous iterations showed that unwanted electroplating occurred under the photoresist, leaking under from the pads, after longer electroplating time. Removing unwanted copper avoids possible leaked electroplating.
\\

A 100ml electroplating bath is developed using the recipe as outlined by Ewing et al. \cite{ewing}, which was suggested to electroplate CoPt onto Silicon substrates. The bath ingredients are shown in Table \ref{tab:table1}. A few modifications to this recipe including the cobalt and ammonium citrate dissolved in DI water prior to adding to the mix, and using a sonicator over a stirrer improved the incorporation of these to the platinum compound.

\begin{table}
\centering

\begin{tabular}{lcr}

Chemical &Concentration (mol/L)\\
\hline
Dinitrodiammine-Platinum & 0.0500\\
Cobalt Sulfamate Hydrate  & 0.0515  \\
Ammonium Citrate & 0.1000 \\
\end{tabular}
\caption{\label{tab:table1}The ingredients and concentrations used in the electroplating bath. The bath had a 100ml volume total during the electroplating process.}
\end{table}

    
    
    

The bath was heated to 50\textdegree C during the electrodeposition, and the pH was frequently checked to ensure that the pH remained at 7. To maintain this pH level NaOH or H\textsubscript{2}SO\textsubscript{4} was added in small amounts, respectively.

The sample is held by clipping the electrical contact on a pad that remains out of the bath and that connects to all of the pads via the copper trace, acting as the cathode. A parallel clip holds a platinum rod and mesh (99.99$\%$ purity) that acts as an anode. The anode needs to be cleaned every 20 minutes by dipping it in 10$\%$ hydrochloric acid (HCl) to remove any impurities caused during the plating process. Therefore, the plating took place in 20 minute intervals, where the plating progress could be assessed. This anode was approximately 2cm away from the PCB during plating.
The source meter was programmed to run the plating for the 20 minute intervals and recorded the voltage every 30 seconds during this time. In general, if there is a sudden drop in the voltage this indicates that the micromagnet has cracked, leaked, or delaminated.
The applied current was decided by the current density; a variety of current densities, ranging from 50mA/cm\textsuperscript{2} to 200mA/cm\textsuperscript{2}, were trialled over a range of plating times from 2 – 4 hours.

The samples were annealed in a tube furnace set to 600\textdegree C for 30 minutes, in a flow of argon gas. The magnets were allowed to slowly cool to room temperature before removal. 
\\

\section{Modelling of Magnetic Particles in Microfluidic Platform with Embedded Magnets}
A multiphysics model is developed in COMSOL to analyse the mechanical behaviour of magnetic nanoparticles in fluidic environment while exposed to the field from integrated magnets . The model incorporate a microfluidic channel, with magnetic nanoparticles flowing in water through the channel. A magnet with CoPt properties is then added next to this structure, allowing the magnetic field to trap the particles. 

The Particle Tracing module runs based on Newton's second law of motion,

\begin{equation}
    \frac{d}{dt} \left( m_p \frac{d {q}}{dt} \right) = {F}_t
    \label{eq:placeholder_label}
\end{equation} 

where F\textsubscript{t} is the total force on the particles, m\textsubscript{p} is the mass of the particles, and q is the particle position. The total force on the particles is: 
\begin{equation}
 F_t = F\textsubscript{m} + F\textsubscript{g} + F\textsubscript{d} + F\textsubscript{B} + F\textsubscript{L}  
\end{equation}

where F\textsubscript{m}, F\textsubscript{g}, F\textsubscript{d}, F\textsubscript{B}, and F\textsubscript{L} are the magnetophoretic, gravitational, drag, Brownian, and lift forces, respectively \cite{ma2009hemispheres}.
\\

Of these, the magnetophoretic, gravity, and drag forces are the dominating in this scenario. The magnetophoretic force is described by:

\begin{equation}
    \textbf{F}\textsubscript{m} = 2  \pi r\textsubscript{p}\textsuperscript{3} \mu\textsubscript{0} \mu \textsubscript{r} K \nabla |\textbf{H}|\textsuperscript{2}
\end{equation}

where \textbf{H} is the magnetic field, $\mu$\textsubscript{r} is the fluid relative permeability, $\mu$\textsubscript{r,p} is the particle relative permeability, and K is defined as:

\begin{equation}
    K = \frac{\mu \textsubscript{r,p}-\mu \textsubscript{r}}{\mu \textsubscript{r,p}+2\mu \textsubscript{r}}
\end{equation}

The magnetic field \textbf{H} relates to the remanence by: 
\begin{equation}
    \textbf{{B}} = \mu_0 \mu_{rec} \textbf{{H}} + \textbf{{B}}_r
\end{equation}

\begin{figure}
\centering
    \includegraphics[width=0.65\linewidth]{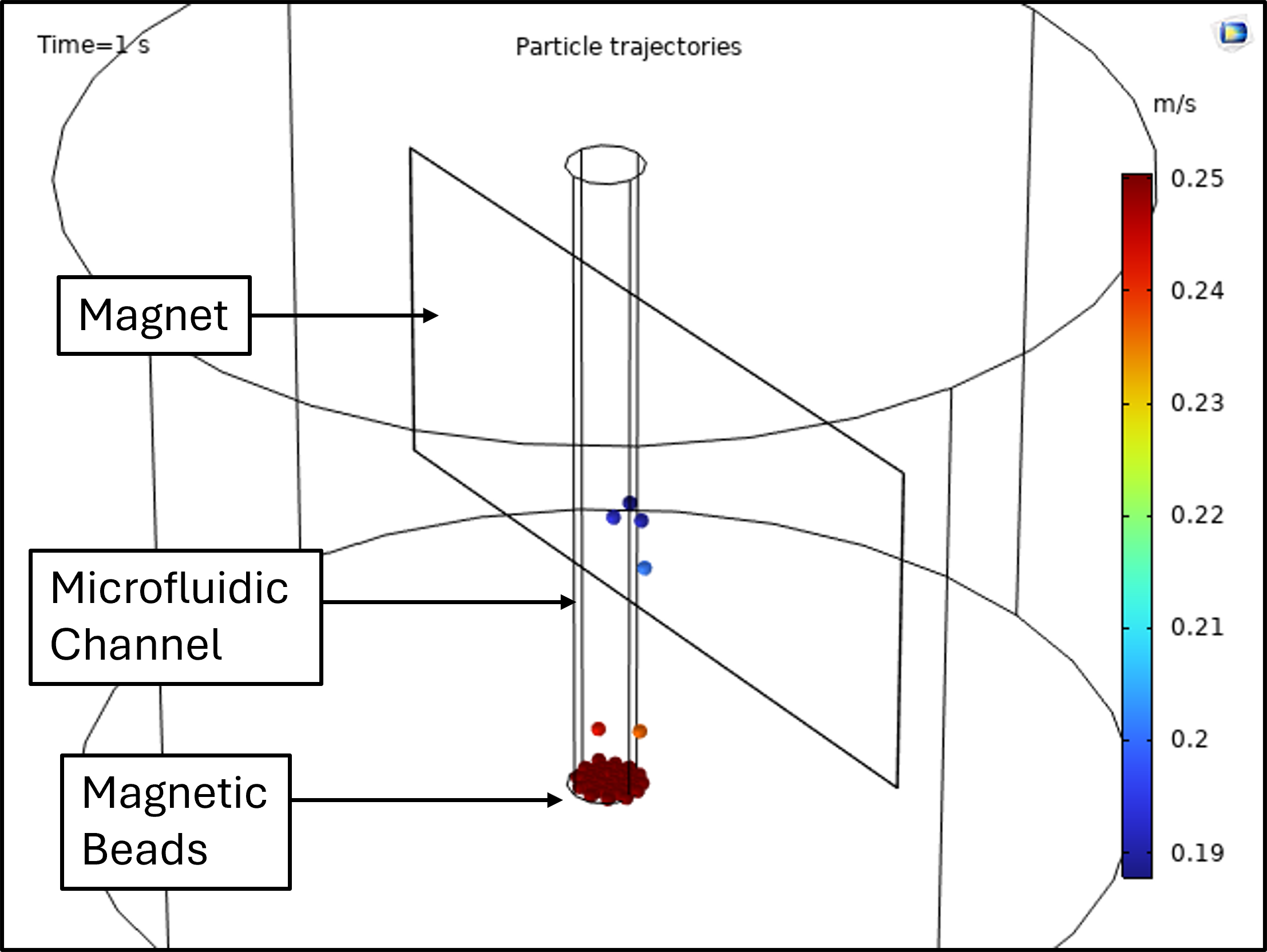}
    \vline
    \includegraphics[width=0.65\linewidth]{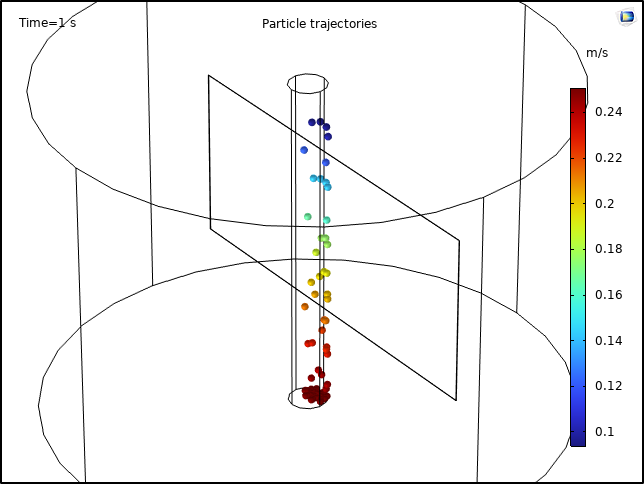}
    \vline
    \includegraphics[width=0.65\linewidth]{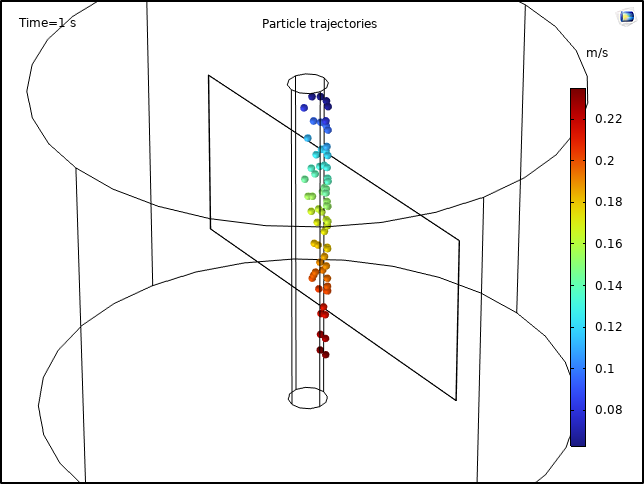}
    \caption{COMSOL simulations of MNPs in water being trapped by magnets with different magnetic field strengths. Top: 500mT. Middle: 1T. Bottom: 1.5T.}
    \label{fig:COMSOL}
\end{figure}

Different remanent fields were applied to this set-up to determine the necessary field to trap some, most, or all particles. The results are illustrated in Figure \ref{fig:COMSOL}.

The modelling uses a 2mm x 4.5mm  magnet, with a thickness of 6$\mu$m. This is to provide the closest match to experimental results. The easy axis is parallel to the out-of-plane field.
The magnet with 0.5T magnetisation traps some particles, while 1T traps the majority. Approximately 1.4T is the required field to trap all particles. 

\begin{figure}
    \includegraphics[width=1\linewidth]{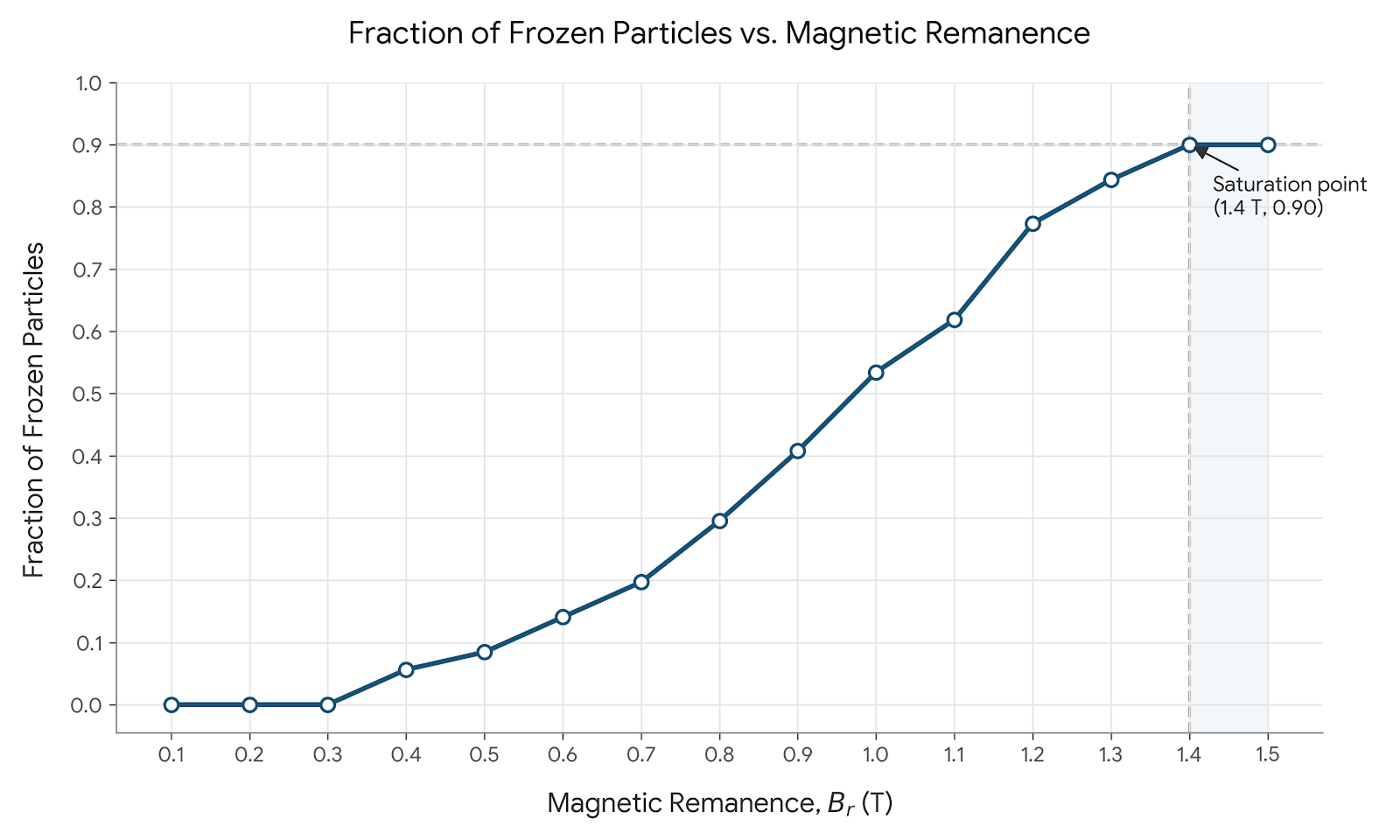}\\
    \caption{COMSOL model of percentage of particles trapped against remanence of CoPt magnet. saturation of particles trapped at 1.4T.}
    \label{fig:sweep}
\end{figure}

This is further investigated by running a parametric sweep of magnetic remanence of embedded magnets to quantify the percentage of nanoparticles released in the channel versus the trapped particles. 
The results shown in Figure  \ref{fig:sweep} confirms the non-linear increase in the percentage of trapped particles versus the magnet strength.
\section{Characterization of Embedded Magnets and Trapping Experiment}



\subsection{Thickness Measurements}
\begin{figure}
    \includegraphics[width=1\linewidth]{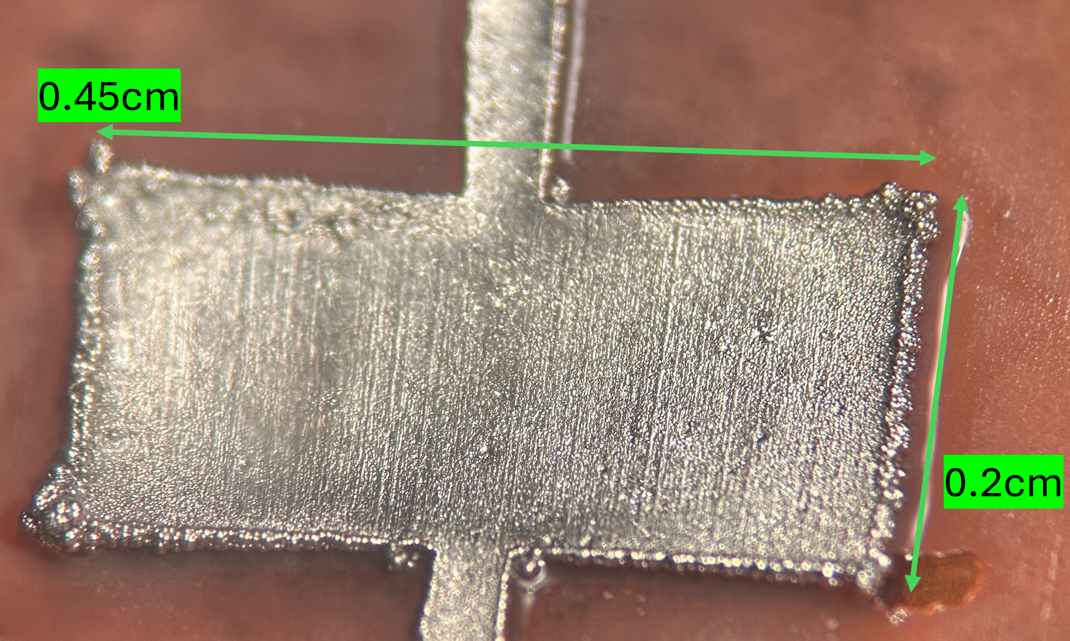}\\
    
    \caption{\label{fig:microscope} Tilted optical microscope images of an electroplated CoPt micromagnet on the Cu-on-Kapton substrate. The dimensions have been included.}
    
\end{figure}

\begin{figure}
\centering
    \includegraphics[width=0.9\linewidth]{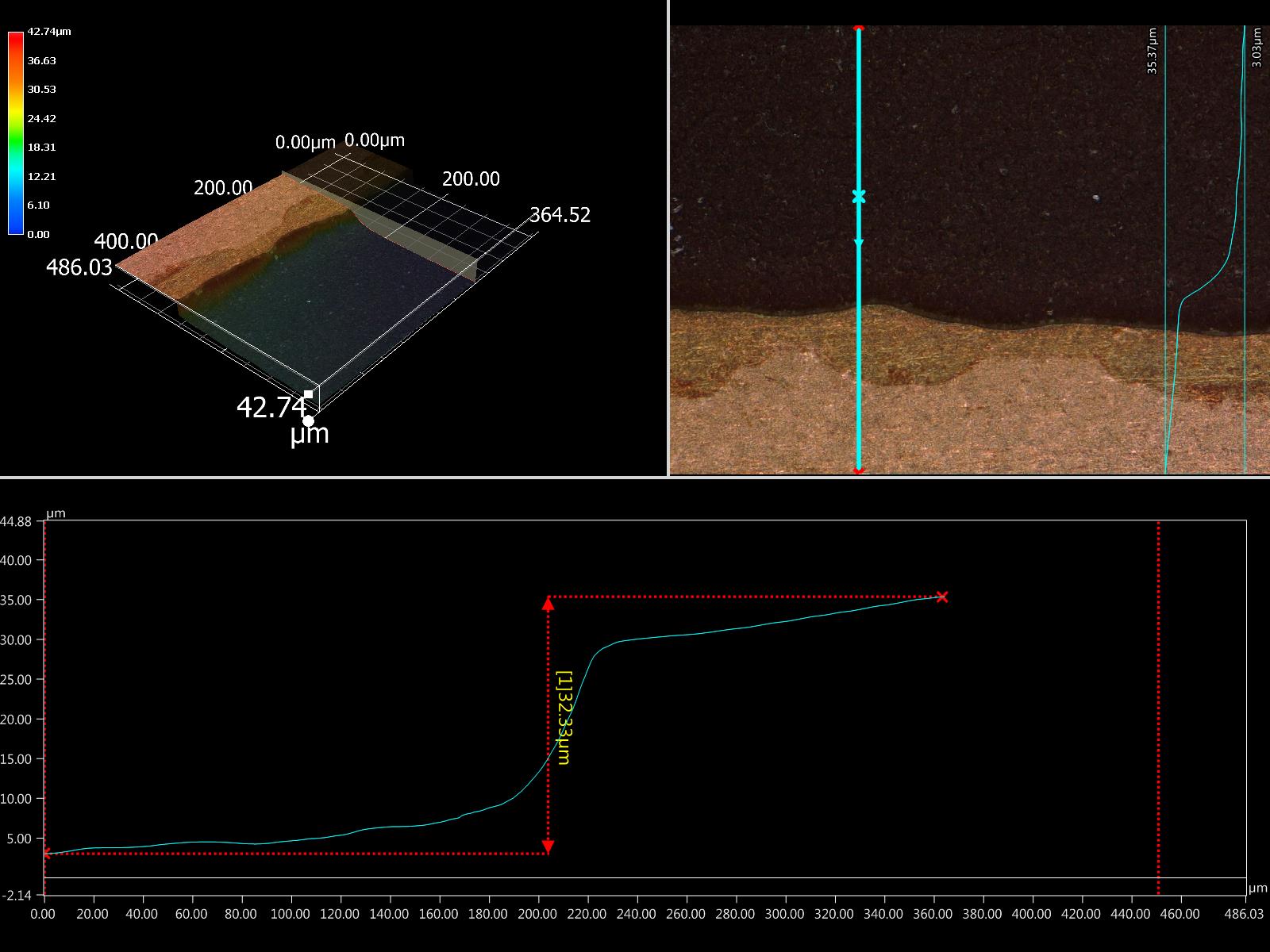}\\
    \includegraphics[width=0.9\linewidth]{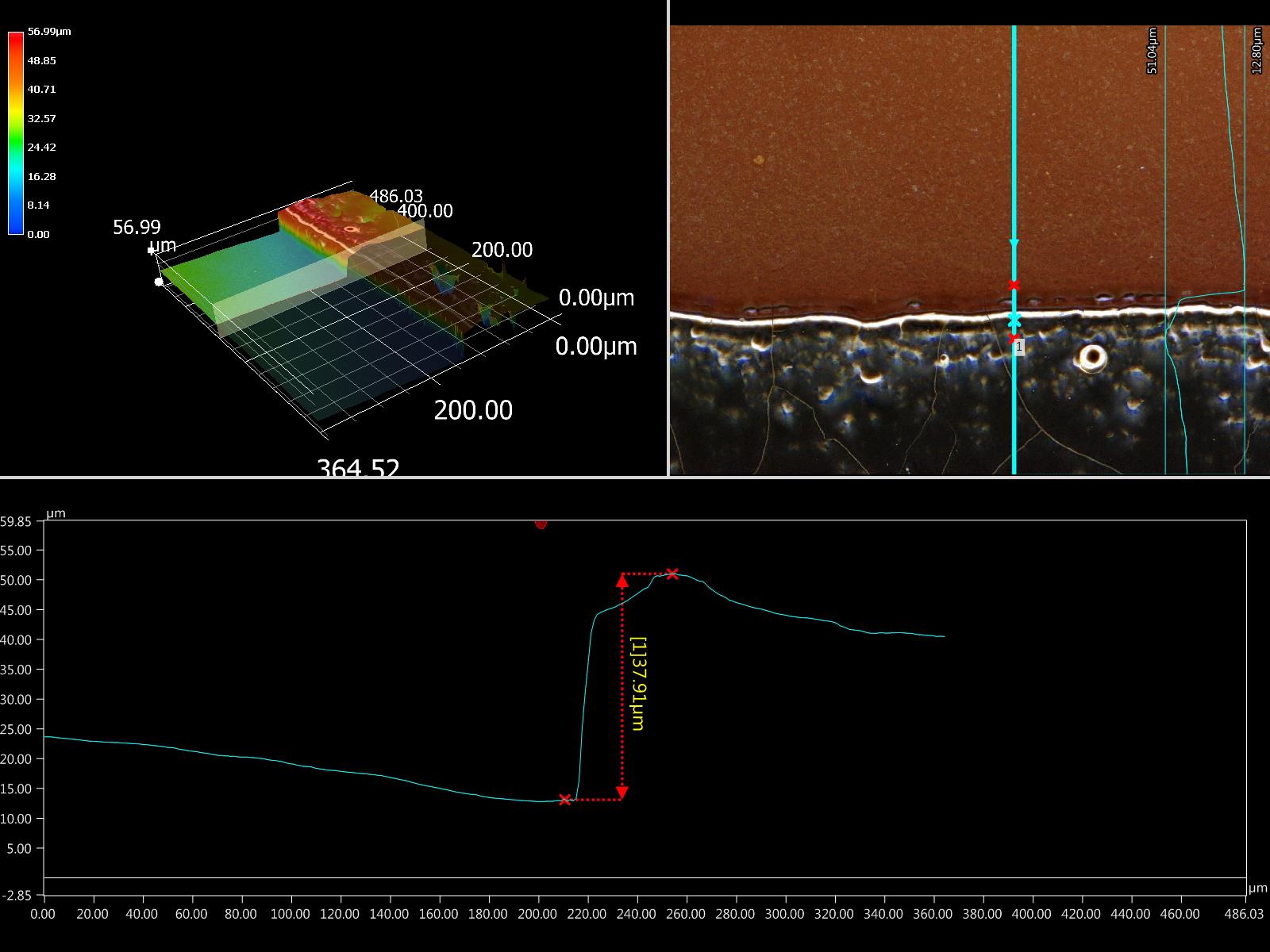}
    
    \caption{\label{fig:thickness} Keyence 3D microscope images of the copper-on-Kapton substrate (top) and the electroplated CoPt on the copper-on-Kapton substrate (bottom). For each image, the top left shows a 3D recreation of the area being investigated, top right shows a microscope image of the area, and the bottom shows a graph of the thickness results based on the blue line seen in the top right image.}
    
\end{figure}

Figure \ref{fig:microscope} provides a slanted optical microscope image of an electroplated pad. The magnets were electroplated for 2 hours at a current density of 100mA/cm\textsuperscript{2}.
To determine the thickness of the magnets, a Keyence 3D Microscope was used. The thickness of the copper on the substrate was first measured, and then this value is removed from the value of the CoPt. The images from the microscope, and the corresponding 3D recreation and profiles, can be seen in Figure \ref{fig:thickness}. The thickness was determined to be around 6.0 $\pm$ 0.2 $\mu$m. 
\\

\subsection{Magnetic Properties}
Before annealing, CoPt showed soft magnetic properties with weak remanence, hence incapable of trapping magnetic beads. However, after the annealing process and consequent phase changes, the hard magnetic properties are developed including highly irreversible ferromagnetic properties with strong remanence; these magnets are now capable of trapping magnetic beads in microfluidic channels.
\\

\begin{figure}
\centering
    \includegraphics[width=1\linewidth]{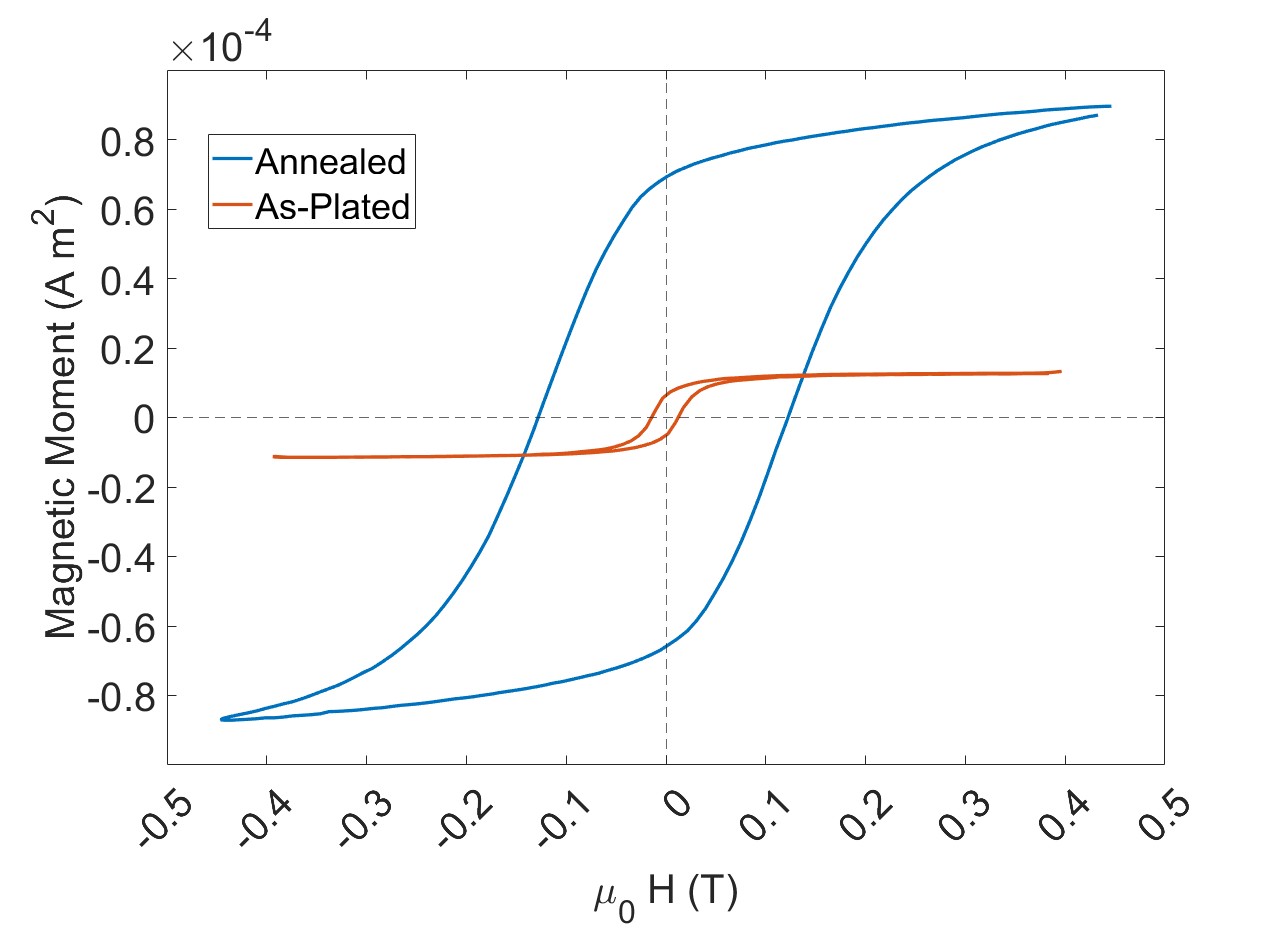}\\
    \includegraphics[width=1\linewidth]{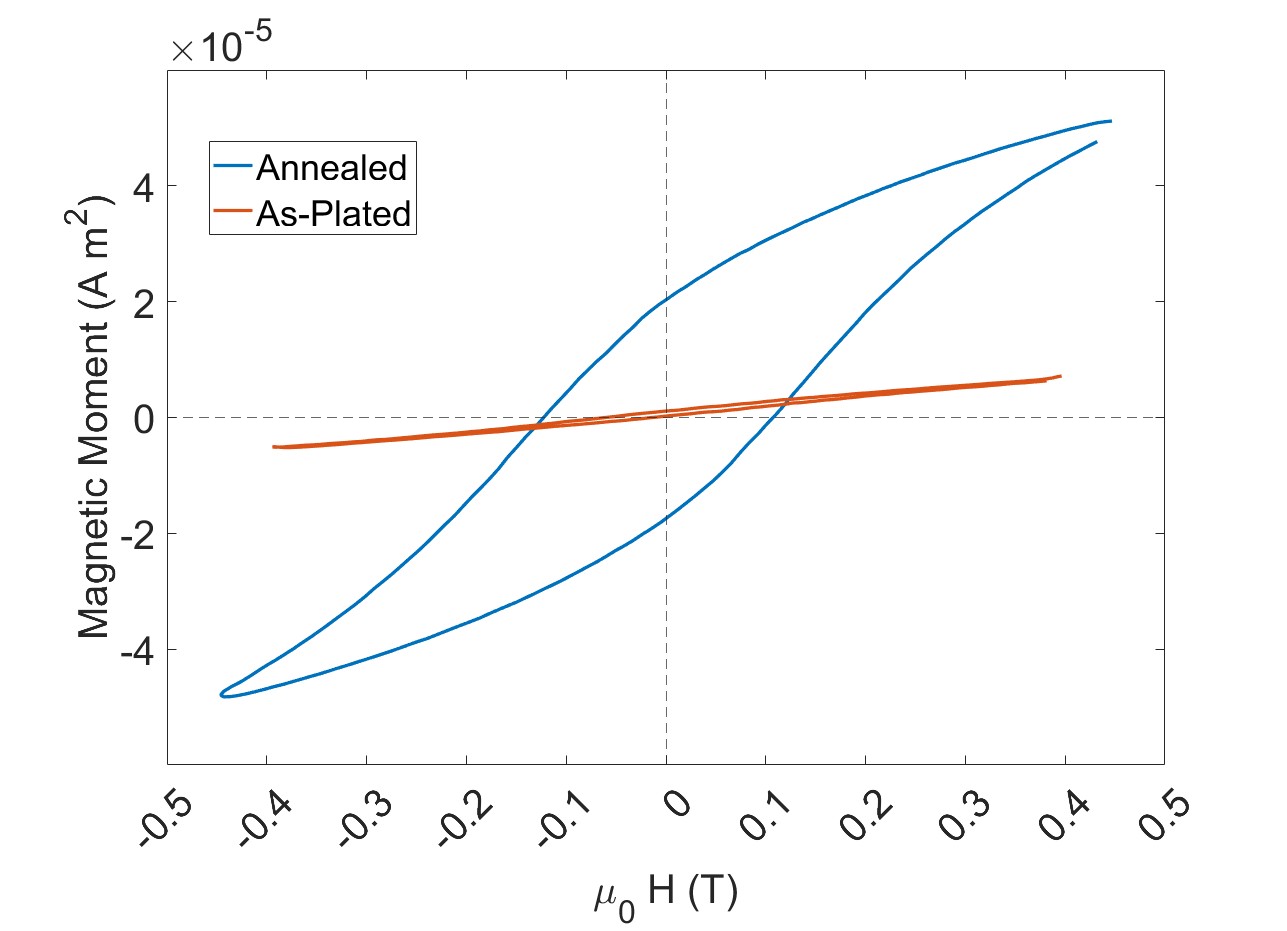}\\

    \caption{\label{fig:vsm}VSM results for the CoPt magnetic sample, before and after annealing. These are demonstrated by the orange and blue lines, respectively. Top: In-plane results (field applied perpendicular to longest side of rectangle). Bottom: Out-of-plane results.}
    
\end{figure}

The magnetic properties of the CoPt magnets were evaluated using a Vibrating Sample Magnetometer (VSM). These were tested after plating and before annealing, and again after annealing, to show the impact of annealing.
The measured results are shown in Figure \ref{fig:vsm}. The hysteresis curves before and after annealing show that the annealing has developed significantly stronger magnetic properties, with the in-plane remanence increasing from 0.24T to 1.4T, which is about six times increase in magnetic strength.
\\

The out-of-plane results also showed a large increase, from 5.7mT to 470mT. The in-plane results are much stronger, and the easy magnetisation axis is parallel with the substrate. 

\subsection{Crystalline Structure}
\begin{figure}
\centering
    \includegraphics[width=1\linewidth]{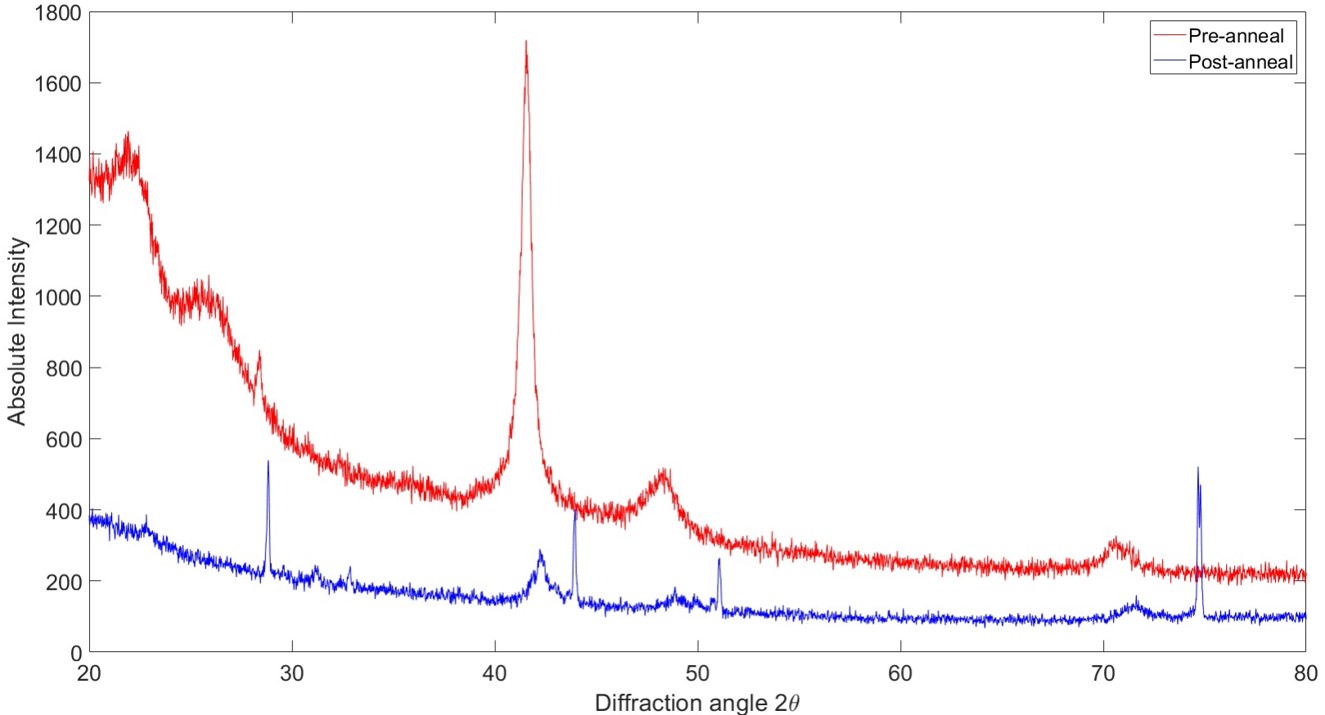}\\
        
    \includegraphics[width=1\linewidth]{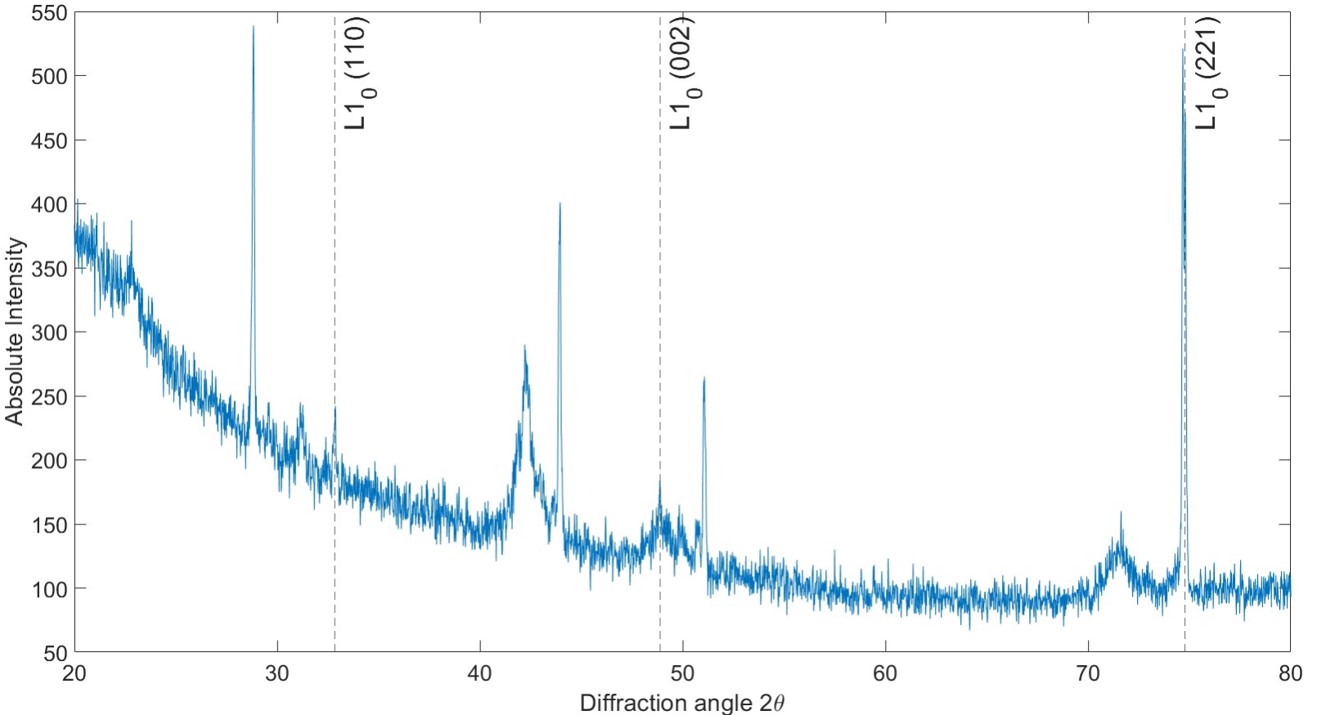}\\

    \caption{\label{fig:xrd}Top: XRD spectra of a pre- and post- annealed sample. Bottom: XRD spectrum of the annealed sample. Some of the important peaks seen only in the L1\textsubscript{0} phase are indexed.}
    
\end{figure}

For further clarification that the magnets had transformed from the disordered A1 phase to the crystalline L1\textsubscript{0} phase, X-ray diffraction (XRD) was used to investigate the material structure. Similarly to VSM, this was run before and after annealing, for comparison between the two states. XRD results demonstrate changes in crystalline structure following the annealing process. 
The XRD analysis shown in Figure \ref{fig:xrd} demonstrates that there is a change in the as-plated versus the annealed samples. In the top graph, blue line of the annealed sample shows several peaks that are not seen in the other data, and is missing the largest peak shown by the other sample. The bottom graph in Figure 4 shows the annealed data in more detail, with some of the L1\textsubscript{0} peaks indexed. 
The lattice parameters \textit{a} and \textit{c} were calculated to be 0.3858nm and 0.3727nm, respectively. The axial ratio a/c therefore was 0.966. This agrees well with previous literature values \cite{thick}.
\\

\subsection{Elemental Composition}

Energy-dispersive x-ray (EDX) analysis was used to determine the elemental composition of the magnet. For an ideal magnet, an equiatomic 50:50 ratio of Co:Pt is desired.
\\

This analysis, set to an accelerating voltage of 10kV, revealed that the CoPt contained 45.9 $\pm$ 1.2 \% Pt, giving a very close to 50:50 ratio of Co:Pt. 
\\

\subsection{Trapping Magnetic Nanoparticles}

\begin{figure}
\centering
    \includegraphics[width=1\linewidth]{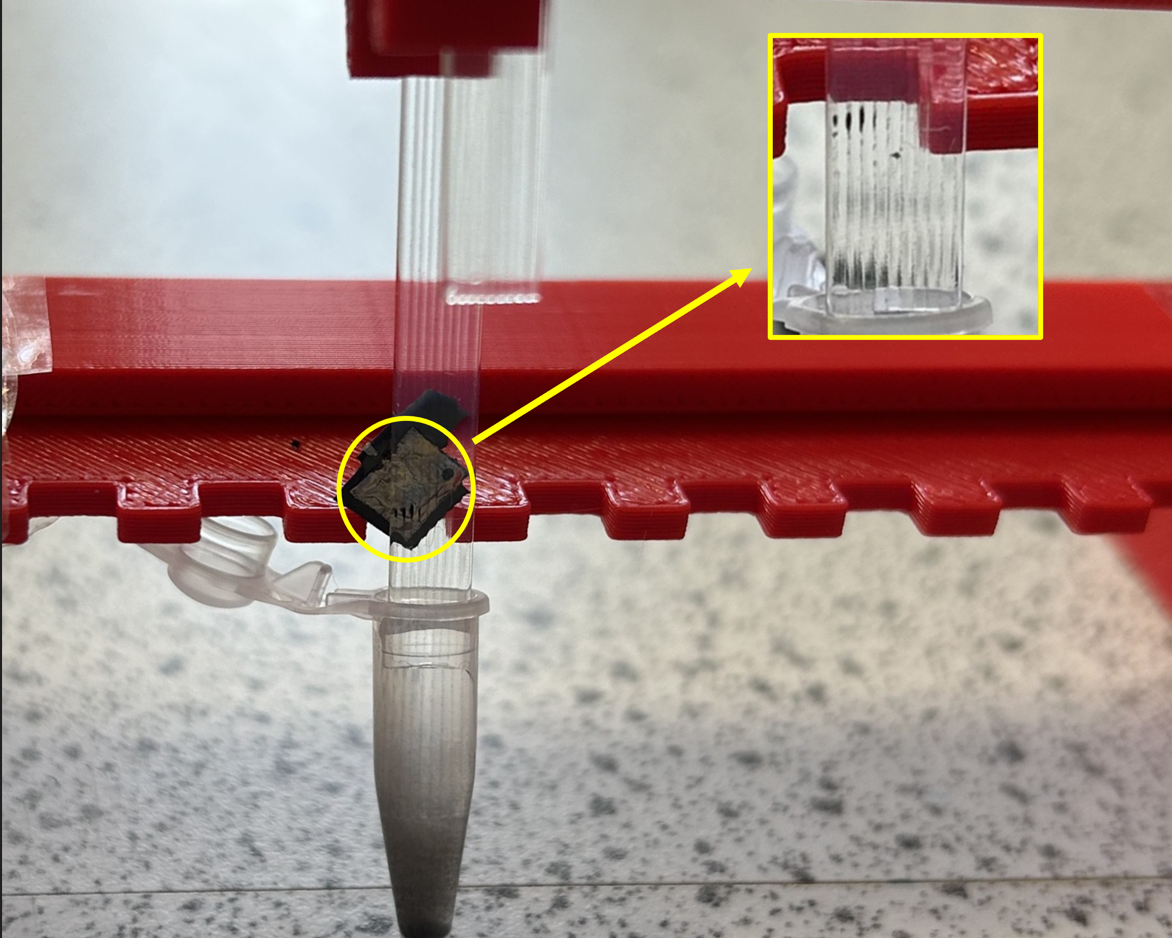}\\

    \caption{\label{fig:beads}CoPt magnet on PCB successfully trapping magnetic nanoparticles as they flow through microchannels. The inset shows the trapped beads remaining in the channels after the removal of the magnet.}
    
\end{figure}

The proposed scalable magnetic platform on PCB was then coupled with microfluidic channels. The magnets plated were successfully used to trap magnetic beads (MagaZorb Reagent) in microfluidic channels. Using a siphon system, as outlined by Ianniello et al. the top of the strip is placed in a mixture of 20µl of magnetic beads and 920µl of distilled water, then curves over the holder, and down into more distilled water to complete the siphon. The magnet is placed alongside the length of the strip, to trap the magnetic particles as they descend through the microchannels, as demonstrated in Figure \ref{fig:beads}. The set-up used here has been inspired by the set-up used in some DNA isolation procedures \cite{ianniello2025dna}.
\\

The magnet, prepared by magnetising to magnetic saturation perpendicular to the easy axis, can be seen successfully trapping some of the beads, and there were no macroparticles that developed and blocked the channel, which has been seen previously with stronger bulk magnets. 
\\

\section{Conclusions}
The MNP trapping demonstrated experimentally agrees well with the expected outcomes as determined by the FEA modelling in COMSOL Multiphysics. The 500mT simulation shows some bead trapping, which matches what is seen experimentally with an out-of-plane remanence of 470mT. 
\\

The process outlined here has successfully created a scalable permanent magnet suitable for use within biosensors. The magnets were electroplated onto a flexible copper-on-Kapton substrate, and successfully annealed to demonstrate crystalline lattice structure and ferromagnetic properties, increasing the coercivity nearly tenfold.
\\

Electroplated micromagnets on a PCB substrate have many potential applications. The micromagnetic lab-on-PCB system described here has been proven to be capable of trapping magnetic beads often used in DNA amplification and cell sorting procedures. 

\section*{Acknowledgements}
We wish to acknowledge Dr Siva Sivaraya for his support, and Crescenzo Ianniello for his expertise. This research has been supported by EPSRC funding EP/S019960/1.





\bibliography{example}





\end{document}